\documentclass[conference]{IEEEtran}

\usepackage{tikz}
\usepackage{amsmath}

\usepackage[procnumbered,ruled,vlined,linesnumbered,noend]{algorithm2e}
\usepackage{hyperref}

\usepackage{xcolor}

\usepackage{listings}
\usepackage{parcolumns}
\usepackage{caption}
\usepackage{enumitem}
\usepackage[font=small,labelfont=bf]{caption}
\usepackage{multirow}
\usepackage{array}
\usepackage{subcaption}

\usepackage{booktabs}
\usepackage{tabularx}
\usepackage{makecell}
\usepackage{multirow}

\usepackage{xcolor,colortbl}

\RequirePackage{totpages}

\usepackage{amssymb}
\usepackage{pifont}

\newcommand*\cmark[0]{$\surd$}
\newcommand{\xmark}{\color{red}{\ding{55}}}%

\newcommand{\lang}{Parasol\xspace}

\newcommand{\note}[1]{{\color{red} #1}}
\newcommand{\todo}[1]{{\color{red}[TODO: #1]}}
\newcommand{\mary}[1]{\textit{\textcolor{green}{[mary]: #1}}}

\newcommand{\dpw}[1]{\textit{\textcolor{orange}{[dpw]: #1}}}
\newcommand{\dl}[1]{\textit{\textcolor{violet}{[dl: #1]}}}
\newcommand{\jsonch}[1]{\textit{\textcolor{teal}{[jsonch]: #1}}}

\newcommand{\ignore}[1]{}
\newcommand{\OMIT}[1]{}

\definecolor{purple}{RGB}{200, 0, 100}
\lstdefinelanguage{lucid4all}
{keywords=[1]{else, if, int, global, Array, entry, exit, event, handle, const, hash, update, generate, update_complex, memop, extern, module, fun, bool, void},
 otherkeywords={}, 
sensitive=true, 
basicstyle=\ttfamily,
keywordstyle=\color{darkgreen}\ttfamily,
keywordstyle=[1]\ttfamily\bfseries\color{black},
comment=[l]{//},
morecomment=[s]{/*}{*/},
keywords=[2]{symbolic, for},
keywordstyle=[2]\ttfamily\bfseries\color{purple},
commentstyle=\color{blue}\ttfamily, 
stringstyle=\ttfamily, 
identifierstyle=\ttfamily,
alsoletter={0,1,2,3,4,5,6,7,8,9}
}
\lstdefinelanguage{p4all}
{keywords=[1]{else, if, register, bit, action, hash, read, write, apply},
 otherkeywords={}, 
sensitive=true, 
basicstyle=\ttfamily,
keywordstyle=\color{darkgreen}\ttfamily,
keywordstyle=[1]\ttfamily\bfseries\color{black},
comment=[l]{//},
morecomment=[s]{/*}{*/},
keywords=[2]{symbolic, for},
keywordstyle=[2]\ttfamily\bfseries\color{purple},
commentstyle=\color{blue}\ttfamily, 
stringstyle=\ttfamily, 
identifierstyle=\ttfamily,
alsoletter={0,1,2,3,4,5,6,7,8,9}
}
\newenvironment{packeditemize}{\begin{list}{$\bullet$}{\setlength{\itemsep}{0.2pt}\addtolength{\labelwidth}{0pt}\setlength{\leftmargin}{\labelwidth}\setlength{\listparindent}{\parindent}\setlength{\parsep}{1pt}\setlength{\topsep}{0pt}}}{\end{list}}

\begin{document}


\title{Automated Optimization of Parameterized\\ Data-Plane Programs with \lang}
%

\author{\IEEEauthorblockN{Mary Hogan\IEEEauthorrefmark{1},
Devon Loehr\IEEEauthorrefmark{1},
John Sonchack\IEEEauthorrefmark{1},
Shir Landau Feibish\IEEEauthorrefmark{2},
Jennifer Rexford\IEEEauthorrefmark{1},
David Walker\IEEEauthorrefmark{1}}
\IEEEauthorblockA{\IEEEauthorrefmark{1}
Princeton University}
\IEEEauthorblockA{\IEEEauthorrefmark{2}The Open University of Israel}
\IEEEauthorblockA{\{mh43, dloehr, jsonch\}@princeton.edu, shirlf@openu.ac.il, \{jrex, dpw\}@princeton.edu}
}
\maketitle

\maketitle
\thispagestyle{empty}

\begin{abstract}
%
Programmable data planes allow for sophisticated applications that give operators the power to customize the functionality of their networks.
Deploying these applications, however, often requires tedious and burdensome optimization of their layout and design, in which programmers must manually write, compile, and test an implementation, adjust the design, and repeat. 
In this paper we present \lang, a framework that allows programmers to define general, parameterized network algorithms and automatically optimize their various parameters. The parameters 
of a \lang program 
can represent a wide variety of implementation decisions, and may be optimized for arbitrary, high-level objectives defined by the programmer. Furthermore, optimization may be tailored to particular environments by providing a representative sample of traffic.
We show how we implement the \lang framework, which consists of a {\it sketching language} for writing parameterized programs, and a simulation-based {\it optimizer} for testing different parameter settings. 
We evaluate \lang by implementing a suite of ten data-plane applications, and find that \lang produces a solution with comparable performance to hand-optimized P4 code within a two-hour time budget.


\end{abstract}

\maketitle
\thispagestyle{empty}

\section{Introduction}
\label{sec:intro}

The advent of programmable data planes has provided network operators the ability to customize the low-level behavior of network switches, paving the way for advanced applications that run inside the network itself. These applications include distributed firewalls, load balancers, sophisticated telemetry, mechanisms for distributed coordination, and application accelerators like in-network caches. Each of these data-plane programs leverages the specialized packet-processing hardware of the switch to run at line rate.

This power comes at a price: sophisticated data-plane programs are notoriously difficult to write and optimize. Thanks to the complexity of switch hardware, programs written in domain-specific languages like P4~\cite{p4} often take hours and even days to compile~\cite{chipmunk}, or fail to compile at all~\cite{chipmunk,lyra,lucid,p4all, sketchguide}, as they cannot fit their data structures into the limited memory and computation constraints available on a chip. And compilation is just the first step. Getting a data-plane program to perform \emph{well} requires making many algorithm-level design choices---thresholds, memory allocation, which data structures to use, etc. Furthermore, these decisions can vary based on the target, performance objective, or expected workload, and are beyond the scope of traditional compilers~\cite{lyra,lucid} and semantic-preserving program optimizers~\cite{chipmunk,lyra, sketchguide}.

Ultimately, today, optimizing a data-plane program is a manual process that requires deep knowledge of the different ways in which its algorithms can be changed, the program's operating environment, and how these factors relate to expected performance. In this paper, we develop a general framework to automate algorithm-level optimization of data-plane programs. 
We argue that such a framework must incorporate three properties, detailed below, to give programmers the ability to express general, parameterized programs. 
We detail how existing systems fall short in~\S\ref{sec:background}.

{\bf High-level performance objectives.}
Application parameters often control low-level details like resource usage, but applications are typically developed to optimize some high-level criteria: accuracy of measurements, effectiveness of sampled distributions relative to ground truth distributions, and bandwidth used are just a few other ways to evaluate data-plane algorithms. An optimization framework should allow programmers to easily express application performance goals.

{\bf ``Program flex.''} 
Data-plane algorithms can be tweaked in so many ways
that affect their performance: the rate at which active probes are emitted in a telemetry application, the choice and size of data structures to use in an in-network cache, the threshold at which to declare a heavy hitter, or the criterion to use for failure detection, to name just a few. As such, a framework should allow programmers to express these design decisions naturally, as parameters in the program. 


{\bf Environmental input.}
The performance of a data-plane application is a product of its environment. A program tailored to one workload may not perform well if used in a different setting. For example, if a programmer wanted to express a property such as ``optimize cache hit rate,'' this would only be possible if the optimization framework considered the expected workload, because hit rate depends on the distribution of requests in the network.

Developing a framework which supports high-level objective functions, program flex, and environment-aware optimizations is challenging because of the trade-off between an abstraction's expressibility and the complexity of the optimization. Expressible abstractions with higher program flex allow programmers to express more general programs, by enabling parameters that can represent a broader range of program components. This inherently makes the optimization more difficult, as the program can have any number of parameters that affect its performance, and it can be nearly impossible to develop objective functions to capture every parameter in a program. 
On the other hand, while limiting the flexibility simplifies the optimization, it also restricts the types of parameters that can be expressed.

{\bf Enter \lang.} \lang is a novel, general framework for synthesizing data-plane programs that fulfills all of the framework goals. \lang consists of two parts: a sketching language and an optimizer. The sketching language is an extension of Lucid~\cite{lucid}, a high-level, event-based data-plane programming language. \lang programmers write \emph{sketches}~\cite{sketch}, 
which are normal programs with several ``holes''. These holes represent the \emph{parameters} of the program; each is an undefined value that will be filled in by the optimizer. Parameters in \lang are highly flexible; they can control just about any aspect of the implementation.
This might include memory layout, decision thresholds, measurement intervals, or even a choice between data structures.
 
The optimizer uses an iterative search algorithm to automatically optimize the parameters according to a user-defined performance objective. These objectives come in two parts. The first part measures arbitrary aspects of a program executing in simulation mode over an example traffic trace. The second part computes an arbitrary, user-defined score based on those measurements. Both parts are implemented in Python rather than the more limited languages of switch data planes. Hence, users can express essentially unlimited optimization criteria---the main constraint is the fidelity of the simulation environment to reality. The optimizer simulates the program's behavior on traffic traces drawn from a particular network, allowing more tailored optimizations than would be possible from relying solely on static, workload-independent quantities such as switch memory resources and architecture.

In summary, \lang is a new data-plane sketching language and optimization framework with the following features:
\begin{packeditemize}
    \item \textbf{Flexible objectives:} \lang's optimization algorithm can optimize for a wide variety of \emph{high-level} metrics such as hit rate or measurement accuracy.
    \item \textbf{Flexible programs:} The parameters of a \lang program may control many  properties, including probe generation frequency, algorithmic choices, memory layout, data-structure selection, or threshold values.
    \item \textbf{Flexible environments:} \lang programmers may tailor their optimization to particular network environments by providing representative traffic traces.
\end{packeditemize}

We evaluate \lang by fully implementing and optimizing ten different data-plane programs, with various parameters and objective functions.
Our experiments found that the \lang optimizer completed a  simulation iteration in approximately eight minutes on average (with an average trace size of two million packets), and all applications could be optimized with a time budget of two hours (i.e., with fifteen iterations on  average). 
The solutions produced by the optimizer not only complied with hardware resource constraints, but were comparable in performance to hand-optimized P4 code.

\section{An Illustrative Example}
\label{sec:background}

\begin{figure}[t]
\includegraphics[width=0.87\linewidth]{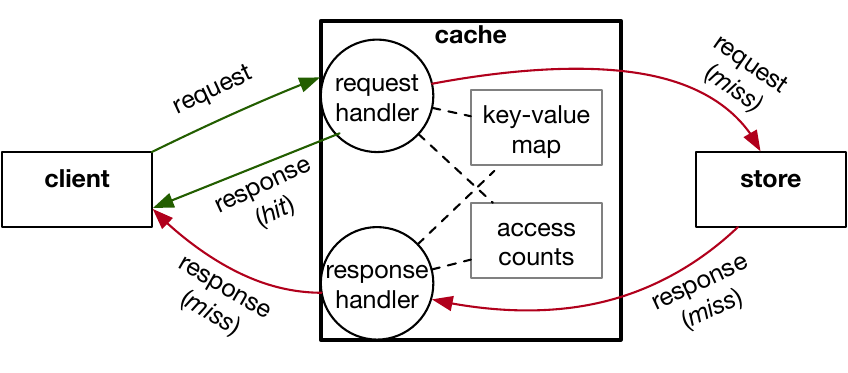}
\caption {Motivating example: an in-network cache. }
\label{fig:cacheexample} 
\end{figure}

\begin{figure}[!t]
\centering
\small
\setlength{\tabcolsep}{2pt}
\begin{tabularx}{\linewidth}{ l X } 
\toprule
\textbf{Param} & \textbf{Description} \\
\midrule
$C_m$ & Number of columns / hashes in multi-hash table (MHT). \\ 
$R_m$ & Number of rows (cells per hash) in MHT. \\
$C_c$ & Number of columns / hashes in count-min sketch (CMS). \\
$R_c$ & Number of rows in CMS. \\
$T_t$ & Timeout threshold for cache. \\
$T_r$ & Replacement threshold for CMS. \\
$P$ & Use precision in place of MHT + CMS. \\
\bottomrule
        \end{tabularx}
\caption{Parameters of the data-plane cache.}
\label{tab:cacheparameters}
\end{figure}


Before describing \lang in detail, we provide a motivating example application that one might wish to deploy in a programmable network: a load-balancing cache, inspired by NetCache~\cite{netcache}. The structure of the cache is illustrated in Figure~\ref{fig:cacheexample}. The cache reduces load on storage servers by directly serving requests for the most popular keys, and forwarding only cache misses to the servers.

The cache operates by storing key/value pairs in a hash table on a switch. When a request arrives, the switch first checks to see if the key is in the table; if it is, the switch simply retrieves the value and sends it back to the requester. Otherwise, the switch forwards the request to the appropriate storage server. When the response arrives, the switch forwards it to the client and optionally caches the entry. 

To maximize efficiency, the cache should serve requests for the most popular keys. Because popularity may change over time, the switch dynamically updates its cache to remove less popular keys in favor of more popular ones. To enable this, the switch tracks statistics about the popularity of keys \emph{not} stored in the cache using a second data structure: a compact, approximate counter (e.g., a count-min sketch (CMS)).

{\bf Parameters.}
%
The high-level description of the cache algorithm is quite simple, but to implement it, a programmer must make numerous low-level decisions.
How much memory should be allocated to the hash table vs. the counter?
When should we replace cached keys?
How should we represent that counter---using a CMS, or something else?
Is a popularity counter even the best eviction algorithm for the cache to use? Perhaps it would be better to use Precision~\cite{precision}, a hash table that probabilistically replaces cached keys upon collision, where more popular items are less likely to be replaced.

Clearly, there are many ways to implement a cache. If we imagine a program that describes all the implementations of a cache that a programmer can imagine, then each of the design questions corresponds to a parameter in that program. Figure \ref{tab:cacheparameters} provides a non-exhaustive list of the parameters in a cache. 

Existing data-plane optimization systems do not give programmers the flexibility to express all of these different design choices as parameters. 
Some, like P$^2$GO~\cite{p2go}, do not give programmers the opportunity to express what can be changed in a program. It applies three fixed optimizations: it can merge tables, remove dependencies, or move processing to the control plane. It cannot modify high-level design choices that affect the precision of an application. 
Similarly, traditional compilers like GCC~\cite{gcc} cannot support the type of parameterization necessary for algorithm-level optimization of data-plane programs---they only optimize for a fixed set of resources (e.g., wall clock time). 

A system like P4All~\cite{p4all} or SketchGuide~\cite{sketchguide} adds a little more flexibility by allowing data structures to be resized. Still, data-plane algorithms could be tweaked in so many other ways that affect their performance. No tool to date allows users to write programs with so much flex, let alone automatically optimize them.

{\bf Performance.}
Parameter value decisions are not simply details --- they can have significant performance implications. For example, a larger hash table can cache more keys, but reduces the memory available to the counter and, in turn, its accuracy. A too-small timeout means that moderately popular keys will get frequently evicted and re-added, while a too-large timeout can result in less popular keys staying in the cache for far too long.

Fundamentally, these trade-offs exist because programmable switches have extremely limited resources that are shared across all data structures on the switch. 
As a result, it is particularly difficult to figure out precisely what effect different decisions will have on the program's behavior.

In contrast, the \emph{desired} behavior of a data-plane cache is easy to define---it should maximize hit rate. This behavior is equally easy to measure, by simply monitoring the switch in question and recording whether each incoming packet is a hit or a miss. 
However, most systems to date focus on optimizing simple on-switch resources~\cite{p2go, clickinc, sketchguide, chipmunk}. Common optimization criteria include memory footprint, number of stages, or ALU usage. 
While optimizing memory layout of these data structures is important, the high-level objective is actually to maximize cache \emph{hit rate}. No tool to date has the ability to specify objectives at such a high level of abstraction.

Additionally, some systems, like P4All, require objectives to be defined as a function of a program's parameters. For the data-plane cache, though, hit rate is not easy to \emph{predict} from the values of the parameters, let alone model analytically. 
It would be very difficult (likely impossible) to derive a closed-form equation that relates the cache's hit rates and parameters, which precludes us from using a system like P4All.


{\bf Traffic dependence.} There is another wrinkle: the hit rate of the cache does not depend solely on the parameters, but also on the network. 
As a result, systems like Chipmunk~\cite{chipmunk} and P4All are limited because they have no access to traffic traces.
While other systems, such as P$^2$GO and SketchGuide, \emph{do} provide access to such data, but they have neither the flex nor the range of objectives to exploit that information to its fullest potential.
P$^2$GO, for instance, optimizes by cutting out program components that are unnecessary for processing a particular traffic trace, but this carries some risk if traffic not present in the trace shows up in the live network.

The importance of optimizing for the expected workload can be illustrated by comparing cache performance across workloads and configurations. The hit rate of a cache depends on which keys are in the cache, which is determined not only by how large the data structures are but also the choice of that data structure (CMS vs. Precision) and the timeout threshold. Certain parameters can have a large range of potential values (e.g., timeout could range from milliseconds to seconds to even longer). The subset of that range that performs well in practice can be quite small, and existing systems such as P4All do not allow us to express parameters such as timeouts, making it easy to pick suboptimal values. 

We found that if a programmer chooses poor parameter values, hit rate for a skewed workload could be as low as 56\%, in contrast to the parameter values that \lang found, which had a hit rate of 93\%. For a uniform workload the hit rate with poor parameters plunged to 11\%, while \lang managed a hit rate of 28\%. 
One might worry that \lang is achieving its better hit rates by overfitting to its input trace; this is a concern for any framework that relies on a particular input. We discuss how to prevent overfitting in detail in \S\ref{subsec:tradeoffs}.

\section{Sketching Language}
\label{sec:language}




The first component of Parasol is a \emph{sketching language} that allows users to write parameterized programs. This language is an extension of Lucid~\cite{lucid}, a high-level data-plane programming language built atop P4. Lucid uses C-like syntax to provide an \emph{event-based} view of the network, in which incoming packets are represented as \emph{events}. When a packet arrives at the switch, the event's \emph{handler} is executed. Handlers run directly on switch hardware, and may read and modify header values and register arrays, as well as drop, create, and forward packets. Lucid provides two backends: a simulation-based interpreter and a compiler to P4.

We chose Lucid as the basis of our tool for two reasons. First, as a high-level language, it provides useful abstractions for representing the numerous decisions programmers must make during implementation. Second, Lucid provides an interpreter that can simulate a program's behavior without compiling it. The interpreter runs a network-wide simulation, and can be run on different input traces, allowing the same program to be optimized for different traffic profiles with no additional user effort.

To implement Parasol, we add four new features to Lucid. First, we add symbolic values (à la P4All~\cite{p4all}) to represent the parameters of a program that should be optimized. Second, we add a way to select between two different data structures based on a symbolic value. Finally, we add a foreign function interface that allows the user to take arbitrary measurements of the network during simulation. Figure \ref{fig:cachecode} shows a pared-down example implementation of a data-plane cache that we use to demonstrate these extensions. Parts of the program that do not relate to \lang's extensions have been omitted, including the hash table storing the key/value pairs.

\begin{figure}[t]
\centering
\begin{minipage}{.45\textwidth}
\begin{lstlisting}[language=lucid4all,basicstyle=\fontsize{8}{10}\selectfont\ttfamily]
symbolic bool useCms;
symbolic int trackerSize;
symbolic int cmsThresh;
symbolic int timeout;

module CMS : {
  type t = ...;
  fun t create(int size) {...}
  fun int getCount(int key) {...}
  fun bool decideIfAdding(int key) {
    return (getCount(key) > cmsThresh);   } }
module Precision : {
  type t = ...;
  fun t create(int size) = {...};
  fun int getCount(int key) {...}
  fun bool decideIfAdding(int key) {...} }
  
module KeyTracker=CMS if useCms else Precision;

global KeyTracker.t tracker = 
    KeyTracker.create(memSize); 
    
extern logHits(bool found);

event request(int key) {
  int cachedKey =   // Hash key and return
  int cachedTime =  // what's stored at that
  int cachedValue = // index in the hashtable
  
  bool found = (key == cachedKey);
  int timeDiff = Sys.time() - cachedTime;
  bool expired = timeDiff > timeout;
  
  logHits(found);
  if (found) 
    { generate response(cachedValue); } 
  else if (expired) 
    { AddKeyToCache(key); }
  else {
    bool add = KeyTracker.decideIfAdding(key);
    if (add) { AddKeyToCache(key); }  } }
\end{lstlisting}
\captionof{figure}{A demonstrative implementation of a data-plane cache in \lang. Parts of the code not containing novel elements have been truncated or omitted entirely. }
\label{fig:cachecode}
\end{minipage}
\end{figure}

{\bf Symbolic values.}
Symbolic values in \lang function as placeholders that may take on any value of the given type. Each is later replaced with a concrete value, supplied during the compilation/optimization process. Once declared, a symbolic is used in the same way as a compile-time constant. 

The program in Figure \ref{fig:cachecode} contains four symbolic values. The boolean \texttt{useCms} determines if the program should use a CMS or Precision data structure, and the integer \texttt{trackerSize} determines how much memory is allocated to that structure. If a CMS is used, \texttt{cmsThresh} determines the threshold for adding new keys to the cache. Finally, \texttt{timeout} determines when keys in the cache are considered expired.

{\bf Selecting data structures.}
Lucid provides a standard \emph{module} system for representing data structures. Each module contains definitions for zero or more types, functions, and events. In Figure \ref{fig:cachecode}, the CMS and Precision modules both contain definitions for a type \texttt{t} --- the type \texttt{CMS.t} represents a count-min-sketch structure, while the type \texttt{Precision.t} represents a Precision data structure. They also contain functions for initializing those structures, and functions for deciding when to add a particular key to the cache.

Although \texttt{CMS} and \texttt{Precision} are the actual modules, they are not referenced anywhere else in the program. Instead, the rest of the program uses the \texttt{KeyTracker} module, which is an alias for either \texttt{CMS} or \texttt{Precision}, depending on the symbolic value \texttt{useCms}. The program may then simply call the function \texttt{KeyTracker.create} to initialize the tracker\footnote{The \texttt{tracker} variable is annotated as \textbf{global} to indicate that it is a persistent structure stored in register arrays.}, and similarly use the function \texttt{KeyTracker.DecideIfAddingKey} to determine if a key should be added to the cache.

\lang's extension to the Lucid type checker makes sure \texttt{CMS} and \texttt{Precision} contain exactly the same declarations (i.e., implement the same interface), which allows the program to use \texttt{KeyTracker} safely while remaining oblivious to implementation-level differences between \texttt{CMS} and \texttt{Precision}. If the modules differed, the programmer could instead create wrapper modules to ensure they present the same interface.

{\bf Foreign function interface.}
Our final extension to Lucid lets a programmer instrument their code with calls to external measurement functions that are executed by the \lang simulator, but removed from the final compiled program. In Figure \ref{fig:cachecode}, 
the extern \texttt{logHits} is a function implemented in Python by the programmer, which counts the number of cache hits and misses while the \lang simulator is running. Each time a cache lookup is performed, \texttt{logHits} is called to record whether the lookup was a hit or a miss. 
After completing a simulation,
the \lang optimizer uses these measurements evaluate the program's effectiveness. 
\OMIT{
For example, in the data-plane cache we have: 
\begin{lstlisting}[language=lucid4all,numbers=none, basicstyle=\footnotesize]
extern count_cache_hits(bool found);
fun int cache_get(int key) {
    int val, bool found = HashTable.get(key);
    count_cache_hits(found);
}
\end{lstlisting}
}

\lang permits only extern functions that have no return value, but does not impose any requirements on what can be passed as a parameter to these functions. Since externs also cannot modify any Lucid program state, this means they can be safely elided during compilation.


\OMIT{
\subsection{Program Sketches}
\lang programs are \emph{sketches}~\cite{sketch} in which parameters, such as the cache parameters in Figure~\ref{tab:cacheparameters}, are declared by the programmer, but automatically assigned values via an iterative search procedure implemented by the framework. The \lang language itself is an extension of Lucid~\cite{lucid}, a high-level, event-based data-plane programming language.

\dpw{change to symbolic values everywhere}
\paragraph{Symbolic values.}  We extend Lucid with 
\emph{symbolic values}---these are ``place holders'' that may take
on any value of the given type.  Concrete values to replace the symbolic
ones will be supplied during the compilation/synthesis process.  
Our data-plane cache program declares three symbolic values. 

\begin{lstlisting}[language=lucid4all,numbers=none, basicstyle=\footnotesize]
symbolic int hash_tbls;
symbolic int hash_rows;
symbolic int timeout;

global Hashtable.t<hash_rows>[hash_tbls] tables;

event clean_index(int i, int j) {
    Hashtable.remove_if_stale(tables[i], j, timeout); }
\end{lstlisting}
\jsonch{Maybe we want to show a pared-down implementation of \texttt{remove\_if\_stale}, rather than this \texttt{clean\_index} event? Then we don't even have to say what an event is.}
This part of the program declares three symbolic values: the number of hash tables to include in the cache, the size of each, and how long inactive entries may remain in the cache before being evicted. After a symbolic value is declared, it may be used in exactly the same way as a compile-time constant. A program can use symbolic values to control the geometry of data structures that are allocated at compile time. In the above code, for example, we use \texttt{hash\_tbls} to create a \emph{vector} of hash tables, each of size \texttt{hash\_rows}. \dpw{vector is italicized, as if a definition
will be given but that definition is never supplied. A reader might not know what a vector is...eg: what the difference between a "vector" and a "array" ... why are these people using that word?} The Lucid keyword \texttt{global} indicates that these hash tables are stored in the switch registers, and hence persist across packets. 

A program can also use symbolic values in decision logic that executes at run time. In the above code, the \texttt{timeout} symbolic value is passed as a parameter to the \texttt{remove\_if\_stale} function, which deletes entry \texttt{j} of hash table \texttt{i} if it has not been accessed in the last \texttt{timeout} nanoseconds. In our implementation, \texttt{remove\_if\_stale} is called when the switch receives a \texttt{clean\_index} event, which is generated when a request for an uncached key collides with an expired cached key. \note{we lazily evict on timeouts; if an uncached key hashes to a slot in cache with an expired key, we will always evict that key to insert the uncached key}\dpw{is this note in red to
be communicated to the reader?}

\ignore{
\begin{lstlisting}[language=lucid4all,numbers=none, basicstyle=\footnotesize]
symbolic num_tbls;
symbolic tbl_size;

global Hashtable.t<tbl_size>[num_tbls] tables;

event clean_index(int i, int j) {
    Hashtable.remove_if_stale(tables[i], j, cache_timeout); }
\end{lstlisting}

This program declares three symbolic values: the number of hashtables to include in the cache, the size of each, and how long entries should remain before being cleaned up. It then creates a \emph{vector} of \texttt{num\_tbls} hashtables, each of size \texttt{tbl\_size}. The Lucid keyword \texttt{global} indicates that these hashtables are stored in register arrays, and hence may be modified by packets during execution.

Finally, the program declares an \emph{event} named \texttt{clean\_index}. Events are Lucid's representation of packets, and may be generated either by packets arriving at a switch or by the Lucid program itself. When an event is generated, the corresponding code (its \emph{handler}) is executed; in this case, the \texttt{clean\_index} event calls a hashtable function to empty index \texttt{j} of table \texttt{tables[i]} if it has not been accessed in the last \texttt{cache\_timeout} nanoseconds. \dl{Or whatever unit of time makes sense here}
}

\paragraph{Data-structure selectors.}
Symbolic values allow program represent the choice of data-structure implementation using symbolic boolean values. For example, as we saw in Section~\ref{sec:background}, the cache benefits from a different implementation of the counter data structure depending on the skewness of the key popularity distribution. In \lang, the cache programmer can write:
\begin{lstlisting}[language=lucid4all,numbers=none, basicstyle=\footnotesize]
symbolic bool use_cms;
module KeyTracker = CMS if use_cms else Precision;
\end{lstlisting}

\dpw{added the type bool in the code above. It is not a bad idea to type check the code in the paper using Lucid just to try to catch typos and things.}

Here, \texttt{CMS} and \texttt{Precision} are data-structure \emph{modules}. A module in Lucid is just a collection of datatype, function, and event definitions, together with an \emph{interface} that specifies how the rest of the program can use those definitions. Although \texttt{CMS} and \texttt{Precision} are the actual modules, they are not referenced anywhere else in the program. Instead, the rest of the program uses the \texttt{KeyTracker} module, which is a reference to either \texttt{CMS} or \texttt{Precision}, depending on the symbolic value \texttt{use\_cms}. 

\lang's extension to the Lucid type checker makes sure \texttt{CMS} and \texttt{Precision} have the exact same interfaces, which is what allows the rest of the program to use \texttt{KeyTracker} safely while remaining oblivious to the implementation-level differences between \texttt{CMS} and \texttt{Precision}.

\begin{figure}[t]
\centering
\begin{minipage}{.45\textwidth}
\begin{lstlisting}[language=lucid4all,basicstyle=\fontsize{8}{10}\selectfont\ttfamily]
symbolic bool useCms;
symbolic int trackerSize;
symbolic int cmsThresh;
symbolic int timeout;

module CMS : {
  type t = ...;
  fun t create(int size) {...}
  fun int getCount(int key) {...}
  fun bool decideIfAdding(int key) {
    return (getCount(key) > cmsThresh);   } }
module Precision : {
  type t = ...;
  fun t create(int size) = {...};
  fun int getCount(int key) {...}
  fun bool decideIfAdding(int key) {...} }
  
module KeyTracker=CMS if useCms else Precision;

global KeyTracker.t tracker = 
    KeyTracker.create(memSize); 
    
extern logHits(bool found);

event request(int key) {
  int cachedKey =   // Hash key and return
  int cachedTime =  // what's stored at that
  int cachedValue = // index in the hashtable
  
  bool found = (key == cachedKey);
  int timeDiff = Sys.time() - cachedTime;
  bool expired = timeDiff > timeout;
  
  logHits(found);
  if (found) 
    { generate response(cachedValue); } 
  else if (expired) 
    { AddKeyToCache(key); }
  else {
    bool add = KeyTracker.decideIfAdding(key);
    if (add) { AddKeyToCache(key); }  } }
\end{lstlisting}
\captionof{figure}{A demonstrative implementation of a data-plane cache in \lang. Parts of the code not containing novel elements have been truncated or omitted entirely. }
\label{fig:cachecode}
\end{minipage}
\end{figure}

\paragraph{External measurement functions.} Our final extension to Lucid lets a programmer instrument their code with calls to external measurement functions that are executed by the \lang simulator, but removed from the final compiled program. For example, in the data-plane cache we have: 
\begin{lstlisting}[language=lucid4all,numbers=none, basicstyle=\footnotesize]
extern count_cache_hits(bool found);
fun int cache_get(int key) {
    int val, bool found = HashTable.get(key);
    count_cache_hits(found);
}
\end{lstlisting}

The extern \texttt{count\_cache\_hits} is a function implemented in Python, by the cache programmer, that counts the number of cache hits and misses while the \lang simulator is running. 
After completing the execution of a concrete configuration of the cache, the \lang simulator reports the state left by the extern functions so that it can be used to score the configuration. 

To support externs in Lucid, we added a simple foreign function interface. We extended the syntax and type system to permit extern functions that take integers and booleans as parameters, but have no return values. Since externs have no return values, and cannot modify any Lucid program state, they are safely elided by an extern elimination pass that we added to the Lucid compiler's backend.

}
\section{Optimizing Sketches}
\label{sec:optimization}

\begin{figure}[t]
\centering
\includegraphics[width=0.7\linewidth]{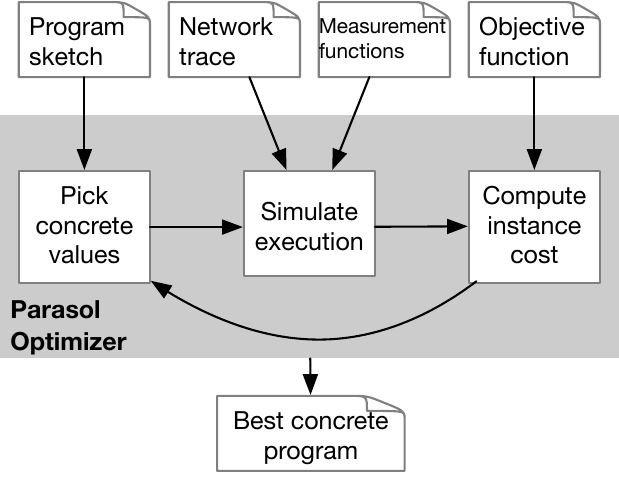}
\caption {Overview of the \lang optimization framework. 
}
\label{fig:overview} 
\end{figure}

The second component of \lang is a framework for automatically optimizing the parameter values of a program sketch; a high-level overview of this framework is provided in Figure \ref{fig:overview}. The programmer provides four inputs: (1) a program sketch, (2) a traffic trace, (3) one or more \emph{measurement functions}, and (4) an \emph{objective function}. The \lang optimizer then finds effective values for the parameters of the program using an iterative search algorithm. In each iteration, the search algorithm selects a concrete value for each symbolic value.
The resulting program is then simulated on the provided traffic trace using the Lucid interpreter.

During simulation, measurements are taken via calls to the measurement functions, using \lang's foreign function interface.
At the end of simulation, the objective function uses these measurements to score the concrete program. The search algorithm then uses the historical series of those
scores to select new concrete values for the next iteration. This process repeats for a set time budget. At the end, the optimizer returns the highest-ranked concrete program that successfully compiles to the underlying hardware.

\subsection{Simulation}
We use a modified version of the Lucid interpreter to model the behavior of \lang programs on a traffic trace. The interpreter simulates the passing of messages between one or more switches in a network, running the appropriate Lucid code when each is received. The simulation includes important switch features such as recirculation and timestamps. To enable execution of \lang programs, we augmented the interpreter to handle symbolic values and foreign functions.

Although the Lucid interpreter models many important aspects of a network, it is not perfect. For example, it provides only a limited model of transmission delay, so properties such as packet reordering are difficult to measure accurately.
However, its limitations are not fundamental; the interpreter could certainly be extended further to accommodate an even wider variety of potential applications.

\subsection{Measurements and Objectives}
\lang optimizes each program according to a programmer-defined objective function, written in Python. The objective can be calculated using any part of the operating environment. Examples of these objective functions include the distribution of flows across paths in a load-balancing application, the rate of collisions in a hash table, and the comparison of a CDF created from run-time measurements to a ground truth CDF. 
The optimizer treats the objective function as a black box; any metric used by the function is acceptable.

Objective and measurement functions are often simple. For our data-plane cache, the goal is to minimize the miss rate (that is, the ratio of cache misses to cache accesses). The functions for measuring and computing miss rate can be defined in just seven lines of Python (Figure~\ref{fig:objcode}). 

\begin{figure}[t]
\centering
\begin{minipage}{.45\textwidth}
\centering
\begin{lstlisting}[language=python, basicstyle=\small]
hits = 0
misses = 0
def logHits(found):
  if found:  hits += 1
  else  :  misses +=1
def objective():
  return misses/(hits+misses)
\end{lstlisting}
\captionof{figure}{Measurement and objective functions in Python for the data-plane cache. }
\label{fig:objcode}
\end{minipage}
\end{figure}

The measurement function \texttt{logHits} is called from the \lang program once per request, as in Figure \ref{fig:cachecode}. The \texttt{objective} function is called by the optimization algorithm at the end of simulation. The global variables \texttt{hits} and \texttt{misses} are maintained in a single instance of the Python interpreter, so their values persist throughout the execution of the program.

{\bf Programmer effort.}
Optimizing a program based on cost and measurement functions greatly reduces programmer effort, compared to prior frameworks that optimize based on closed-form functions. 
Writing a closed-form function to represent the miss rate for a data-plane cache is an arduous task. 
A key is evicted from the cache when there is a hash collision, so the miss rate is influenced by the probability of collisions. 
However, if the key tracker is a CMS, the choice to insert an uncached key after a collision depends on the stored count for that key, and thus, the miss rate also depends on the accuracy of the counts in the CMS. 

The interaction of all these factors is not straightforward - they depend on the workload distribution. Theoretical models would then have to make assumptions about that distribution~\cite{cms}.
\emph{Even if} the programmer goes through the considerable effort of working out a closed-form objective function for a cache, it can only express a theoretical miss rate; the actual rate may be drastically different in practice~\cite{sketchAccuracyIMC}. 

{\bf Comparing with ideal implementations.}
A particularly useful type of measurement is to compare the runtime behavior of a data structure against an idealized implementation. As an example, a data-plane application can produce round-trip time (RTT) samples by matching SYN packets with corresponding SYN-ACKs~\cite{rtt, dart}. When the switch sees a SYN packet, it stores its timestamp in memory, and can compute the RTT when it sees the corresponding SYN-ACK packet. However, if the structure is full, the switch cannot store new SYN packets; as a result, the application can only provide a portion of RTT measurements. During simulation, a measurement function could maintain a Python data structure which does not run out of memory, and compare its results to those of the \lang structure---this provides an easy-to-compute ground truth for how well the \lang program could perform.

\subsection{Search Algorithm}

The final component of the \lang optimizer is the search algorithm itself. The goal of the search algorithm is to find parameter values that minimize the objective function. However, the space of possible solutions can be intractably large.
Doing an exhaustive search is inefficient, and a na\"ive strategy may never discover a compiling solution. 

As a strawman solution, \lang could require users to define the search space by providing bounds on all variables. However, this will almost certainly include a large number of non-compiling solutions, as even experts would have trouble determining the correct bounds. As an example, reasonable bounds on cache with a CMS as the key tracker might be 1-5 cache tables and CMS rows, and cache entries and CMS columns that are less than the amount of memory in a stage. These bounds produce a solution space of 4225 configurations, only 16\% of which compiled to our target switch. 

Alternatively, \lang could use a heuristic to test if each configuration will compile before it is simulated. If the configuration does not compile, \lang can assign it a maximum cost. While this avoids simulating non-compiling configurations, it also reduces the effectiveness of the search strategies, as it does not give any indication of a direction in which to search. 
One could imagine simulating anyway, in the hopes that it will lead us to a compiling configuration, but this is unlikely -- programs using an impossible amount of memory, for example, are likely to perform impossibly well.

In practice, we address this issue by splitting the search algorithm into two phases: preprocessing and simulation.
In the first phase, \lang automatically prunes non-compiling solutions from the search space, without requiring user-defined bounds. 
In the second phase, \lang searches the space of remaining solutions with a user-configurable search algorithm.


{\bf Preprocessing.} In a nutshell, the goal of the preprocessing phase  is to ensure our solutions are making maximal use of the resources on the switch, without using so many that the program fails to compile. Accordingly, during this phase we only consider symbolic values which affect resource allocation.
The resources we consider are memory, pipeline stages, hash units, array accesses, and ALU usage. 
\OMIT{
identify valid parameter configurations that will compile to the target switch. Hence, in this stage, the \lang optimizer only considers parameters that affect the resource usage of the application.
\todo{We identify resource vs non resource variables automatically in program. (we don't do this yet, but we should be able to)}
}
We assume that the program is monotonic with respect to resources --- that is, increasing the value of any symbolic value should not decrease the amount of resources used. In our experience, this is a safe assumption; we note that all of the applications we evaluated satisfied this property. 


The optimizer begins by setting all symbolic values to either a default or user-provided starting value. We then pick a symbolic, and determine an upper bound for it by iteratively increasing only that symbolic's value until we run out of resources. Thanks to monotonicity, the largest value that fits
provides an upper bound for the symbolic. 

We then pick another symbolic and repeat this process; however, this time we find one upper bound for each possible value of the first symbolic. 
We do the same for the next symbolic, and the next, each time finding an upper bound for all valid combinations of previously-processed symbolics. When we finish, we will have enumerated the entire useful search space (i.e., every compiling solution). 

This process, however, grows multiplicatively with the number of parameters. To make it more tractable, we use domain knowledge to set a reasonable default starting value that allows \lang to discover the entire useful search space, without having to compile every solution in that search space. Values that represent memory used per stage are initially set to the max memory available in a stage, and values that contribute to other resources start at 4. 
We choose 4 as a starting value because we found it generalized well to all of our applications, providing a significant reduction in preprocessing time when compared to a starting value of 1. For example, the preprocessing time for caching structure with a Precision key tracker improved from almost 2 hours to only 25 minutes.
{\bf Simulation.} In the second phase of the search algorithm, we perform a configurable search through the pruned space of solutions we created in phase 1. We choose a configuration from phase 1, select values for any non-resource symbolics, and execute the resulting program in the Lucid interpreter. We then score the configuration based on its output, and use a search strategy to select the next configuration to evaluate based on the history of scores. 
%

The \lang optimizer does not rely on any particular search strategy; rather, it is able to accommodate a variety of search algorithms. We provide four built-in search functions for programmers to use---exhaustive search, Nelder-Mead simplex method, simulated annealing, and Bayesian optimization---but \lang also supports any programmer-defined search of the solution space, and is compatible with any technique written in Python (e.g., stochastic gradient descent, genetic algorithms, etc.). 
We choose these strategies because (with the exception of exhaustive) they use the history of scores to efficiently navigate the search space. They also provide a range from simple (exhaustive, Nelder-Mead simplex) to more complex (Bayesian). 
Programmers are free to choose the search algorithm that provides their preferred balance between search time and optimality of the final result.  We evaluate the effectiveness of each of these strategies and analyze how the choice of strategy affects the optimizer in \S\ref{sec:evaluation}.

\subsection{Design Tradeoffs}
\label{subsec:tradeoffs}

\begin{figure}[t!]
\centering
\begin{tabularx}{\linewidth}{l c c}
\toprule
\textbf{Heuristic} & \textbf{Avg compile time} & \textbf{Reduction} \\
\midrule
\textbf{Dataflow graph} & 51s & -- \\ 
\midrule
\textbf{Greedy layout} & 51s & 13\% \\
\midrule
\textbf{Lucid-P4} & 1.5min & 13\% \\
\midrule
\textbf{Full compilation}  & 1.5min & 16\% \\
\bottomrule
\end{tabularx}
\caption{The performance of preprocessing heuristics for a single configuration, averaged over each evaluated application. The greedy layout provides the best balance between performance and accuracy. }
\label{fig:preprocesstime}
\end{figure}

{\bf Accelerating preprocessing.}
The first phase of optimization requires analysis of the resource usage of a program to determine if it will compile. The simplest way to do this would be to actually compile the program; however, compilation can be very slow (the Conquest~\cite{conquest} application took over 13 minutes), and most applications require compiling many configurations (Conquest has a compiling search space of 25 configurations). Instead, we have tested a range of heuristics, with varying trade-offs between performance and accuracy. We provide a detailed description of each heuristic in Appendix~\ref{app:heuristics}.

All three of our heuristics operate by attempting to assign each action in the Lucid program to a stage of the switch's pipeline. The primary distinction between the heuristics is the types of resources they account for during placement. They range in complexity from only considering dependencies between actions (dataflow graph), to modeling every resource except packet header vector (PHV) constraints (compiling from Lucid to P4). We note that heuristics can only underestimate, never overestimate, resource usage. In other words, the solutions that do not compile with a heuristic will also never compile to the target device.

A summary of the performance of our heuristics appears in Figure \ref{fig:preprocesstime}. 
We list the average compile time for each of our evaluated applications and the average reduction in search space size, using the dataflow graph heuristic as the baseline. 
In practice, we have found that the greedy layout heuristic provides the best trade-off between performance and accuracy. We cope with the potential inaccuracy of the heuristic by including a safeguard to ensure that \lang returns a compiling solution. Specifically, we actually compile the highest-ranked configuration at the end of our optimization loop. Should compilation fail, \lang tries the next-highest-ranked, and so on, until one compiles. If none of the tested solutions compile, the system will repeat the optimization process, excluding solutions that did not compile. 

We found that in practice, this rarely happens. 
After manually fixing any PHV errors, the optimal solutions for nine out of the ten applications fit within the target resources. One of the applications (CMS) resulted in ``optimal'' configurations that did not compile. However, the \lang optimizer found a compiling solution that had similar performance. 
{\bf Unrepresentative traces.} Since the \lang optimization framework is simulation-based, it relies on a representative traffic trace. If the actual traffic in the network deviates from the patterns in the trace, the performance of the application may not match the simulated performance. However, because \lang preserves the semantics of the data-plane program, it will never produce unexpected or invalid behavior---its performance may simply be poorer than anticipated.

To mitigate poor performance, programmers can use multiple traffic traces to optimize their application, and use a weighted combination of performance on the traces as the objective function. We show an example with our data-plane cache in \S\ref{subsec:cacheeval}, by optimizing with workloads of different distributions.
Alternatively, if the distribution depends on time of day, the programmer can use traces from peak times, where applications are likely most sensitive to poor performance.

Beyond poor performance, an unrepresentative trace can leave an application vulnerable to attacks when the training trace only contains benign traffic.
To use Parasol for tuning a security system, one needs traces containing the kinds of attacks the application seeks to detect or prevent. Fortunately, \lang users need not acquire and label such traces themselves, as the network security community already goes to great lengths to produce and share traces for the evaluation of their own security systems~\cite{impactrepo}. These traces come from a variety of sources, including cyber defense exercises~\cite{nccdc} and security-oriented testbeds or simulators~\cite{veksler2018simulations, chadha2016cybervan}.


\section{Evaluation}
\label{sec:evaluation}


Our evaluation of \lang addresses its two components:
\begin{packeditemize}
    \item \textbf{Language.} Can \lang express a wide variety of parameters, objective functions, and data-plane applications?
    \item \textbf{Optimizer.} How well do optimized \lang programs perform, and how quickly does \lang find good parameters?
\end{packeditemize}

\noindent
To answer these questions, we used \lang to implement and optimize a suite of ten data-plane applications with respect to representative traffic traces. We chose applications that encompass a wide array of structures (including commonly used structures like sketches and hash tables) and contain a diverse set of parameters and objective functions. Our application and optimizer code is publicly available. \footnote{https://github.com/mhogan26/Parasol}

In the remainder of this section, we discuss each \lang component individually, and finish with two in-depth case studies.
We used three types of traces in our evaluation---the University of Wisconsin Data Center Measurement trace~\cite{uwtrace}, a trace from core Internet routers~\cite{caida}, and synthetic traces for the cache application. Unless otherwise stated, we split a single input trace into a training trace and testing trace (see Figure~\ref{fig:optspeed} for trace sizes).


\begin{figure*}[t]
\begin{tabular}{@{}lcccclcc@{}}
                     & \multicolumn{4}{c}{\textbf{Classes of parameters in application}} & & \multicolumn{2}{c}{\textbf{P4All/S.G.?}}\\
                     \cmidrule{2-5} \cmidrule(l){7-8}
                      & \textbf{mem.}     &  &    \textbf{data struct.}   &        & & & \\
\textbf{Application}  & \textbf{alloc.} &   \textbf{threshold}      & \textbf{choice}  & \textbf{timing} & \textbf{Objective (LoC)} & \textbf{Params} & \textbf{Obj.} \\
Count-min sketch (CMS)                  &  \cmark & & &  & Mean estimate Error (20)& \cmark & \cmark\\
Multi-hash table (MHT)                  &  \cmark & & &  & Collision ratio (11) & \cmark& \cmark\\
Data plane cache (KV~\cite{flightplan}) & \cmark & \cmark & \cmark & \cmark &Miss rate (23)  & \xmark & \xmark\\
RTT monitor (RTT~\cite{rtt})            & \cmark & & & \cmark  & Read success rate (118) & \xmark &\xmark\\
Unbiased RTT (Fridge~\cite{fridge})     & \cmark & \cmark & &  & Max percentile error (88)& \xmark&\cmark\\
Starflow~\cite{starflow}                & \cmark & & &  &    Eviction ratio (17)  & \cmark &\xmark\\
Conquest~\cite{conquest}                &  \cmark & & &  &  F-score (101) & \cmark &\xmark\\
Load balancing (LB~\cite{flowletlb})    &  \cmark &\cmark & & &Error vs. optimal (38)& \xmark&\xmark \\
Precision~\cite{precision}              &  \cmark & & &  & Avg. error for top flows (28) & \cmark &\xmark\\
Stateful Firewall (SFW~\cite{lucid})    &  \cmark & \cmark & & \cmark &Packet overhead (70)& \xmark&\xmark \\
\bottomrule
\end{tabular}
\caption{Applications optimized with \lang, showing which classes of parameters/objective functions were used, and which of them could be expressed in P4All or SketchGuide.}
\label{fig:langeval}
\end{figure*}


\subsection{Language} 

To evaluate the expressiveness of \lang, we implemented applications with multiple classes of parameters and diverse objectives. The right two columns of Figure~\ref{fig:langeval} show the high-level benefit of \lang over prior optimization frameworks: whereas \lang allowed us to fully express the optimization goal of each application (parameters and objective function), P4All and SketchGuide could only express the full optimization goals of 2/10 applications. In the rest of this section, we discuss the ability of Parasol to represent a diversity of both parameters (its ``program flex''), and objective functions.




{\bf Program flex.}
As Figure~\ref{fig:langeval} shows, the Parasol programs we implemented had four general classes of parameters: memory allocation, decision thresholds, choice of data structure, and operation timing. These classes encompassed a diverse range of parameters, including data structure size and the probability of an item being added to a structure.
Parasol's flexible approach allowed it to handle all of them. 
In comparison, P4All and SketchGuide, the only prior frameworks for application-level parameter optimization, could only support parameters from 6/10 of our implemented applications (CMS, MHT, Starflow, Conquest, Precision) as it is impossible to express threshold, timing, or data structure choice parameters in P4All or SketchGuide. 



Even for the examples that \emph{could} potentially be optimized by P4All or SketchGuide, it is easy to imagine slightly more complex variants that would require incompatible parameters. For example, our CMS is a simple implementation with no concept of time intervals---it never resets. Most applications, however, will want to count over intervals, which requires a mechanism to periodically reset or age counters, and a parameter that controls the length of the interval. The addition of that one simple parameter makes the ``deployable'' variant of CMS incompatible with P4All and SketchGuide.

{\bf Objective functions.}
The objective functions for our applications measured a wide variety of high-level properties (Figure~\ref{fig:langeval}). These functions were generally short and simple: on average, each function was approximately 50 lines of Python code. The only requirement for \lang objective functions is that they be expressible in Python. They can include any, all, or none of the parameters in the application, along with any measurements taken during the simulation.

In contrast, existing systems (P4All, SketchGuide) require programmers to supply a closed-form objective function, which specifies exactly how the parameters relate to the final cost. In practice, this can be very difficult, particularly for applications that do not have theoretical guidelines or proven error bounds. This is common, even in research, where many data-plane applications are evaluated empirically, without finding provable theoretical guarantees~\cite{starflow, conquest}.
Furthermore, many systems are composed of multiple components or data structures; writing a closed-form function for those systems requires not just understanding each component individually, but codifying precisely how they interact. 

In our evaluation, we considered an objective function to be expressible in P4All or SketchGuide only if we could find a derivation in existing literature. We consider deriving a closed-form objective function to be beyond the scope of an application developer (and also this paper) as it requires significant theoretical work. 
We required that functions include all the parameters of the applications, but did not require those parameters to be expressible in P4All or SketchGuide. Although functions needed not be for a single component, we note that none of our applications with multiple structures (KV, Starflow, Conquest) had a closed-form function.

With these criteria, we were only able to express three out of our ten objective functions in P4All or SketchGuide. Even so, there is a caveat: functions from the literature typically quantify worst-case performance. These objective functions oftentimes do not provide a realistic idea of how the application performs in practice.
In contrast, \lang objective functions measure actual performance on a sample trace, and are therefore able to optimize for a much broader range of criteria, even when a closed-form error function exists~\cite{sketchguide, sketchlib, sketchAccuracyIMC}.
We compare \lang against a closed-form objective for the unbiased RTT (Fridge) application in detail in \S\ref{subsec:fridgeeval}.



\subsection{Optimization Quality}
\label{subsec:optquality}

We evaluate the quality of \lang's solutions, compared to both hand-optimized systems and an oracle optimizer (described below) 
and analyze the factors that impact it. All experiments in this section are based on a 
two-hour time limit for the dynamic search phase of the \lang optimizer. 

First, we compare the results of optimization with \lang to optimization with an ``oracle''. Whereas the \lang optimizer chooses parameters on a training data set, separate from the testing data, the oracle optimizer chooses parameters by exhaustively searching the testing data set, i.e., it always chooses the optimal parameters. 

\lang found configurations that performed as well as the oracle for 6/10 applications (CMS, MHT, RTT, Starflow, Precision, and SFW). For 3/10 applications (KV, Fridge, Conquest), the relative difference between the objective score of \lang's and the oracle's configuration was under 12\% (i.e., $\frac{|\text{Objective}_\text{oracle} - \text{Objective}_\text{Parasol}|}{\text{Objective}_\text{oracle}}$). For the remaining application, the load balancer (LB), \lang's solution was, in relative terms, 82\% worse than the oracle. However in absolute terms the difference was small: the oracle's configuration performed 1.7\% worse than a perfect load balancer, while \lang's configuration performed 3.1\% worse than a perfect load balancer.

\subsubsection{The \lang preprocessor}

To measure the effect that \lang's preprocessor had on the solution quality, we compared application performance when optimized with and without preprocessing, using the same two-hour time budget for \lang's search phase. When the preprocessor was disabled, we bounded the search space by setting the same initial bounds for all memory allocation variables --- 20 register arrays (e.g., cache tables) and 
the max amount of SRAM per stage for registers (e.g., cache entries per table). In our judgement, this represented a reasonable bound -- high enough to include all compiling solutions for each application without unnecessarily inflating the search space.
Additionally, without the preprocessor, we assigned a predetermined max cost to solutions that did not compile (e.g., 100\% cache miss rate), to avoid simulating them.

\ignore{
\begin{figure}
    \includegraphics[scale=.32]{figures/preprocessed_improvementtest.pdf}
    \caption{\note{fake data --> add firewall, remove cache}; Application performance improvement due to preprocessing. The \% improvement represents the gap in performance on a test trace with solutions found after optimizing with and without preprocessing. The solutions found after preprocessing consistently outperformed those found without.\dl{Should we add "...even though they were present in the search space in both cases", or is that too much?}}
    \label{fig:preprocessimprovement}
\end{figure}
}

Preprocessing consistently improved application performance, especially for applications that had a large search space or used multiple structures that compete for resources (Starflow, KV, Conquest, SFW). In fact, when the cache used CMS as the key tracker, \lang consistently did not find a compiling solution in the time budget without preprocessing. 
\begin{packeditemize}
    \item For Conquest, enabling the preprocessor improved recall from 75\% to 87\%.
    \item For Starflow, the preprocessor improved eviction ratio from 35\% to 15\%. 
    \item For the stateful firewall, the preprocessor improved recirculation and retransmission overhead from 16 kbps to 0.01 kbps.    
\end{packeditemize}

Applications that had a small search space (CMS, MHT, Fridge, LB) did not perform significantly better when preprocessing was enabled. However, even for such applications, preprocessing still has an important benefit: it automatically bounds the search space for the programmer, without the need for them to manually ``guess'' reasonable bounds.

\subsubsection{The \lang searcher}
We found that the effectiveness of \lang's search phase depended on two factors: the search strategy and the quality of the input trace. \lang provides four built-in strategies: exhaustive search, Bayesian, simulated annealing, and Nelder-Mead simplex. We note that all of these strategies (except exhaustive) have hyperparameters that control the learning process. We chose hyperparameter values manually such that strategies produce solutions as good as or near the oracle solutions. We found that we could re-use these values for all applications without negatively affecting solution quality.

\begin{figure}
\centering
    \includegraphics[scale=.32]{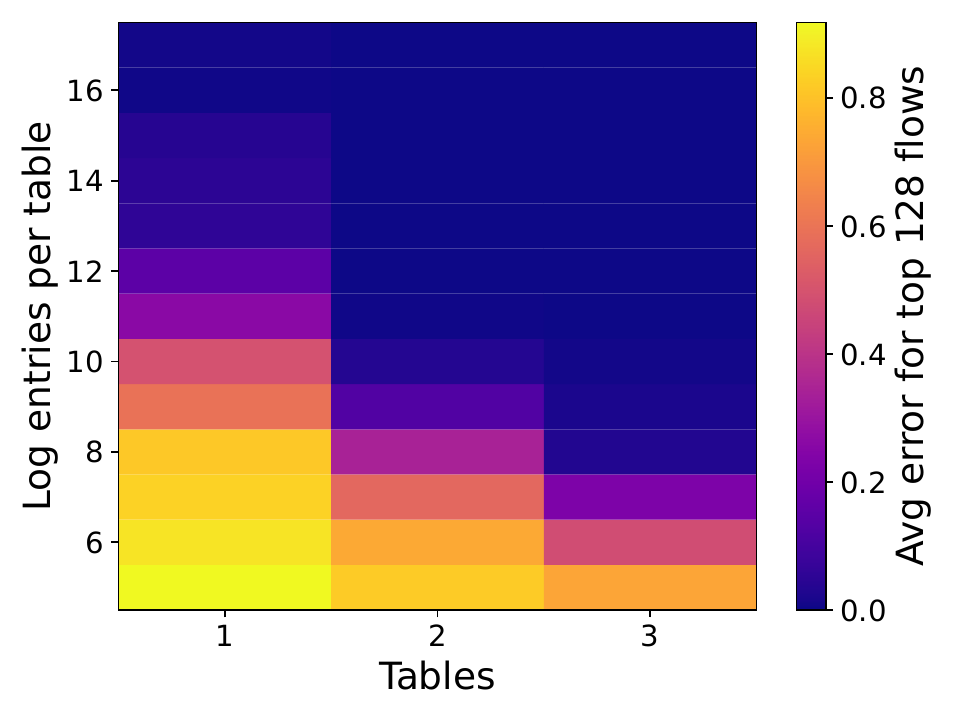}
    \caption{Average error for top 128 flows in the Precision application for different configurations. A darker color represents a lower error. The optimal configuration achieved an error of 0.01\%, and nearly 40\% of the solution space produced an error of less than 1\%.}
    \label{fig:heatmap}
    
\end{figure}

{\bf Search strategy.}
For some applications, the choice of search strategy does not matter because a large portion of the compiling solution space is near-optimal. For example, in the Precision application, over half of the search space after preprocessing contained solutions that produced an average error
of less than 10\% (Figure~\ref{fig:heatmap}), compared to the optimal of less than 1\%. In such cases, the search methods mostly converged to the same configuration or to configurations that had very similar performance. 

For more complex applications, we found that no single search strategy dominated. Because of this, we found that the best strategy was to run multiple strategies in parallel for each application, and choose the best result from among them. Conversely, for applications with a small search space (after preprocessing), we simply used exhaustive search.
We consider a search space to be small if the exhaustive search completed within the two-hour time budget.

{\bf Training trace.}
%
Across all applications, we found that traces with approximately 1 million packets were sufficiently large for \lang to find high quality (i.e., near optimal) configurations. Training trace size mattered more for some applications than others. One large class of applications where training trace size mattered was applications that use hash tables. Here, traces had to be large enough to cause hash collisions; otherwise the differences between configurations are small and it is difficult (or impossible) for \lang's search algorithm to find the best one. For example, the Starflow configurations found by the simplex and Bayesian strategies resulted in similar eviction ratios (12\% and 5\%, respectively) in a small trace of 5000 packets, but had very different errors (46\%, 26\%) with a larger trace of 5 million packets. 

\ignore{
For some applications, it was important that training traces contained distributions representative of the target workload. For example, in the unbiased RTT (fridge) application, \lang found a configuration with an 18\% error in the testing trace, when the training trace had the same distribution \jsonch{[distribution of what??]}. However, \lang used a training trace with a different distribution, the error rose to 40\%. \dl{I don't really follow this paragraph. The first sentence seems obvious (of course the trace has to be representative), and I'm not totally sure I get what the rest of it is getting at. Plus, aren't we automatically representative if we're using a random training/test split?}
}

It was often important that the training trace was representative of the testing trace. For some applications, the search phase was only effective when a trace contained certain network events. For example, the Conquest data structure detects microbursts, and only begins monitoring when one occurs. A trace with no microbursts would produce no meaningful objective, regardless of the configuration. 


Some applications, however, were less sensitive to differences between training traces and target workloads. When testing Starflow on a wide-area network (WAN) trace, we found that \lang was able to find near-optimal solutions using training traces from either a WAN or a datacenter. 

\subsubsection{Comparison to hand-optimized configurations}
We compared the performance of \lang configurations to that of hand-tuned configurations for our three most complex applications: Fridge, Conquest, and Starflow. The hand-tuned configurations come from the applications' original evaluations~\cite{fridge,conquest,starflow}. Our goal is to determine whether \lang can essentially reproduce these results, by finding configurations that perform comparably on a similar workload. 

Appendix~\ref{app:hand_opt} describes the case studies in detail, but at a high level, \lang solutions performed reasonably close to the hand-optimized solutions for all three applications. 
\begin{packeditemize}
    \item For Fridge, \lang found a configuration that achieved a delay estimation error of 28\%, compared with the original evaluation's result of 25\%.
    \item For Conquest, \lang found a configuration with a precision of 97\% and recall of 87\%, compared to the original evaluation which found precision and recall $>$ 90\%, using the same trace. 
    \item For Starflow, \lang found a configuration with an eviction ratio of 15\% in a wide-area workload, which is better than the 18\% eviction ratio reported in the original evaluation of the single configuration that the authors compiled to the Tofino. 
\end{packeditemize}

\begin{figure}[t!]
\centering
\begin{tabularx}{\linewidth}{l l l l}
    \toprule
\textbf{App} & \textbf{Preprocess}& \textbf{Train trace} & \textbf{Test trace} \\
 & \textbf{time} & \textbf{size, time}  & \textbf{size, time} \\
\midrule
CMS & 16s & 500k, 25s & 10M, 12min \\
MHT & 15s & 1M, 47s & 10M, 7min \\
KV, Precision & 25min & 1M, 6min & 5M, 25min\\
KV, CMS & 2hrs & 1M, 2min & 5M, 7min \\
RTT & 23s & 1M, 1min & 3M, 3min \\
Fridge & 3s & 1M, 1min & 3M, 2min \\
Starflow & 1.5hr & 900k, 1min & 5M, 27min \\
Conquest & 15s & 10M, 9min & 10M, 10min \\
LB & 2s & 500k, 16s & 3M, 2min \\
Precision & 32min & 1M, 6min & 18M, 1.7hrs\\
SFW & 30s & 4M, 3min & 11M, 7min \\
\bottomrule
\end{tabularx}
\caption{Runtime of \lang components per application. Preprocess time is the total time to preprocess with the greedy layout heuristic, train/test trace size is the size of the trace in packets, and train/test trace time is the average time to simulate the trace once. }
\label{fig:optspeed}
\end{figure}

\subsection{Optimizer Speed}
The runtime of the Parasol optimizer is application-dependent (shown in Figure~\ref{fig:optspeed}), and has two major components: preprocessing time and search time. Preprocessing time scales with the complexity of the input program and number of parameters, and took between 7 seconds and 1.5 hours. Search time scales primarily with the size of the input trace, and was limited to 2 hours, though many applications required less than that. A single iteration of the training trace took between 16 seconds to 9 minutes, depending on the application.  

Overall, the \lang optimizer took no more than 3.5 hours to find near-optimal settings for any of our applications.
This compares favorably to compiling, testing, and tuning applications by hand: just compiling \emph{one} configuration of a program to a reconfigurable architecture like the Tofino can take hours~\cite{chipmunk} for both research or industrial compilers, because it is a fundamentally hard task~\cite{vass2020compiling}.
As mentioned above, we found three main factors that influenced the overall runtime: application complexity, training set size, and search strategy. 

{\bf Application complexity.}
The optimizer preprocesses each \lang program as a heuristic to check if it will compile to hardware. The preprocessing time depends on the complexity of the program, both in terms of length and number of parameters. Programs with more parameters (e.g., Starflow) took longer than programs with few parameters (e.g., LB).
Figure~\ref{fig:optspeed} lists total preprocessing time for each application.

Complex programs also take longer to simulate. The CMS simulation took about a minute for a 1 million packet trace, while a trace of the same size with Precision took three minutes. 
Precision is more complex because it contains logic for recirculating packets, while the CMS does not recirculate packets. 
The recirculation not only adds complexity to Precision, it also requires the program to process more packets, as recirculated packets must be processed again.


\ignore{
\begin{figure}
\begin{subfigure}{\columnwidth}
    \centering
    \includegraphics[scale=.36]{figures/flowlet_best_per_iter2.pdf}
    \caption{The minimum error in the flowlet load balancing application after each iteration for each search strategy.}
    \label{fig:flowlet_best_per_iter}
\end{subfigure}

\begin{subfigure}{\columnwidth}
    \centering
    \includegraphics[scale=.36]{figures/starflow_strats.pdf}
    \caption{The minimum collision in the starflow application for each search strategy.}
    \label{fig:firewall_best_per_iter}
\end{subfigure}
\caption{Search strategy performance for Bayesian, simulated annealing, and simplex strategy. \todo{make these less ugly/more readable? or just remove them?} \todo{add a vertical line for wall clock time}}
\label{fig:best_per_iter}
\end{figure}
}

{\bf Training set size.} The runtime of \lang's search phase increases roughly linearly with the size of the input training trace, because the search algorithm executes each chosen configuration on the trace. 
Reducing the size of the provided trace can speed up optimization, but many applications require large traces. For example, evaluating the performance of a program that measures heavy hitters (e.g., Precision) requires enough traffic that the trace contains heavy flows.

{\bf Search strategy.}
Search strategies took different amounts of time to converge, depending on the application. We compare search strategies, using the load balancing and Starflow applications, by tracking the best evaluated configuration after \emph{each} iteration. 
All three methods found similarly performing configurations for the load balancer, but the overall search time was much different: Bayesian search took approximately 19 minutes, while simulated annealing and simplex search took only 2 minutes. 
Similarly, for the Starflow application
Bayesian and simulated annealing strategies reached a configurations with similar performance (in 13 and 10 minutes, respectively) while simplex did not find a configuration that produced the best collision rate within the time budget.





\subsection{Case Studies}

\subsubsection{Data-plane caching}
\label{subsec:cacheeval}

To better understand how \lang handles workload dependence and some of the challenges in tuning data-plane applications, we study a conceptually simple in-network cache. We optimize the cache for three different workloads: a highly skewed zipfian (top 10 keys had 58\% of requests), moderately skewed zipfian (top 10 keys had 15\% of requests), and uniform (top 10 keys had .06\% of requests). Training traces contained 1 million requests, and test traces contained 5 million requests.
We limit the cache size to 10K entries. 

We compared three versions of the cache: a variant that uses a count-min sketch to track key popularities (NetCache~\cite{netcache}); one that uses Precision~\cite{precision} to prioritize popular keys; and a very basic hash-addressed array that evicts on collision to avoid the need for packet recirculation.

\begin{figure}[htb]
\centering
\small
\begin{tabular}{lccc}
\toprule
distribution & CMS & Precision & Hash table \\
\midrule
high skew & 0.10 & 0.07 & 0.10 \\
moderate skew & 0.69 & 0.64 & 0.70 \\
uniform & 0.73 & 0.73 & 0.73 \\
\bottomrule
\end{tabular}
\caption{Cache performance with respect to miss rates.}
\label{fig:cachecase1}
\end{figure}

First, as Figure~\ref{fig:cachecase1} shows, all three caches reduce the workload of the backend that they serve. As expected, the caches perform better in more skewed workloads, and the more sophisticated CMS and Precision caches outperform the simple hash table. In particular, the Precision cache is over 30\% more effective than the other caches, in the high skew workload.

\begin{figure}[htb]
\centering
\small
\begin{tabular}{lccc}
\toprule
distribution & CMS & Precision & Hash table \\
\midrule
high skew & 0.5750 & 0.5375 & 0.5500 \\
moderate skew & 1.0175 & 0.8225 & 0.8500 \\
uniform & 1.0475 & 0.8675 & 0.8650 \\
\bottomrule
\end{tabular}
\caption{Cache performance with respect to network workload.}
\label{fig:cachecase2}
\end{figure}

Now consider a network operator with a different objective. Instead of minimizing miss rate, they wish to minimize total network traffic. Assuming the client and server are connected via one hop across the caching switch, a cache miss costs 2X as much as a cache hit, and a recirculated packet costs 0.5X as much as a cache hit. Thus, the objective function is $2*m+h+0.5*r$, where m, h, and r are the percentage of misses, hits, and recirculated packets in a trial. The cache provides benefit whenever the metric is less than 1. 

Figure~\ref{fig:cachecase2} compares the caches with respect to this alternative metric. Somewhat surprisingly, the simple hash table performs \emph{better} than the more sophisticated CMS variant, and is competitive with Precision (even beating it for the uniform workload). It is because the hash table variant does not need to recirculate packets, unlike the others.

This case study highlights how tricky it can be to tune even a conceptually simple data-plane application. The optimizations that at first seem most effective (or are most intuitive) are not necessarily best in every network, or from every perspective. Figuring out what's right for one's network can be challenging, but \lang simplifies this process by lifting the burden of reasoning about how parameter choices can affect performance off of the programmer.

\subsubsection{Fridge}
\label{subsec:fridgeeval}


Sometimes, operators tune their programs with heuristics derived from closed-form equations based on worst-case error bounds. Such heuristics have two problems. First, they are challenging to derive and verify, hence only available for certain data structures whose properties have been theoretically analyzed. Second, closed-form equations do not always give an accurate picture of application performance in practice, because actual traffic distributions can vary significantly from the worst case~\cite{sketchAccuracyIMC} and, for tractability, closed-form equations often ignore factors that matter in practice.

To illustrate this, we compare the performance of the Fridge RTT monitor as optimized by \lang to a version optimized according to a closed-form equation. The Fridge data structure measures RTT by storing requests and matching them with the corresponding response, without sampling bias against samples with large RTTs.
Each request is added to the structure with probability $p$, and a request is removed either when it is matched with a response, or if a new request overwrites it (because of a hash collision). Given a Fridge size $M$ (the number of entries), the authors derive the following formula to set $p$: $\frac{M}{p} = $ number of requests between the request and response with the maximum delay. 

For our evaluation workload, with a Fridge sized at $M = 2^{17}$ (the maximum size for our implementation on the Tofino), the theoretical formula calculated $p=2^{-1}$, which resulted in a maximum percentile error of 31\%. As expected, this was \emph{not} the optimal configuration for this workload. Optimizing with \lang improved the relative performance by 10\%; \lang recommended $p=2^{-5}$, which resulted in an error of 28\%. 

In Service-Level Agreements (SLAs) with ISPs, delay requirements are often specified as a target distribution, or a maximum delay for a certain percentile. It is then essential for ISPs to be able to accurately measure the delay distributions in their networks.
The theoretical formula provides only a worst-case error bound, though, and it is not tailored to the more specific needs of users.
\lang, on the other hand, can easily optimize for users' target SLAs; programmers need only adjust the objective function.
As such, the gap between \lang and the theoretical formula was even more substantial at specific percentiles - for RTT samples in the 50\textsuperscript{th} percentile, the configuration from the theoretical formula produced an error of 15\%, while \lang achieved an error of only 8\%.

Although closed-form equations based on worst-case analysis are important for theoretical rigor, this short case study demonstrates that using them for tuning leaves significant performance on the table. \lang allows network operators to reclaim that potential performance by automatically tuning data structures for different operating environments and performance objectives, while at the same time freeing programmers from the burden of deriving tuning heuristics from worst-case performance bounds.

\section{Related Work}
\label{sec:related}

Researchers have developed a number of 
tools for writing data-plane programs.
Domino~\cite{domino}, Chipmunk~\cite{chipmunk}, ClickINC~\cite{clickinc}, Lucid~\cite{lucid}, Lyra~\cite{lyra}, and O4~\cite{o4} provide new, high-level languages for expressing data-plane programs, each providing abstractions and a compiler targeting one or more architectures. These compilers include optimizations or synthesis techniques to ensure that programs compile. However, if a program cannot fit on a target, it will not compile. 
In the case of ClickINC, the compiler will attempt to place the program on a different device if it cannot be compiled on a switch. They also do not provide environment-specific optimizations, as compilers do not have access to traces. 

There also exist tools for optimizing prewritten data-plane programs. P$^2$GO~\cite{p2go} uses a traffic trace to minimize the resources used by a P4 program by reducing dependencies that do not appear in practice, shrinking tables, and offloading parts of the program to a controller. Cetus~\cite{cetus} uses static analysis to eliminate dependencies between tables. 
P$^2$GO and Cetus either do not provide environment optimizations or risk changing program semantics.
Additionally, Pipeleon~\cite{pipeleon} seeks to optimize the performance of programs deployed on SmartNICs by analyzing runtime performance. In constrast, \lang focuses on resource-constrained programmable devices that cannot be updated at runtime without recompilation.

A third type of tool optimizes by leveraging user domain knowledge. P5~\cite{p5} uses a high-level description of the network's policy to remove spurious dependencies and unused features. P4All~\cite{p4all} and SketchGuide~\cite{sketchguide} allow users to declare flexibly-sized structures and optimize them with a user-provided objective function. 
These tools ask a lot of their users; P5 requires a high-level policy description, and P4All and SketchGuide require a closed-form objective function.

An area of work related to \lang's optimizer is network simulation. Simulators are designed for many objectives, including high fidelity~\cite{ns3}, interactive operation~\cite{mininet}, automatic traffic generation~\cite{mimicnet}, and scalable performance~\cite{maxinet}.  All of these tools complement \lang, and future work will likely involve integrating these tools 
to improve \lang.



\section{Conclusion}\label{sec:conclusion}


The process of writing and deploying a data-plane application that works well is an arduous one, requiring the programmer to undergo a grueling process of compiling, testing, and tweaking to find the best configurations. \lang is a new and flexible framework for writing \emph{parameterized} data-plane programs, and synthesizing effective settings for those parameters. Parameters in \lang can represent a wide variety of high-level implementation decisions, and the \lang optimizer can target a variety of high-level behavioral goals. The optimization process is orders of magnitude faster than modern iterative testing strategies, and incorporates a representative traffic trace to tailor its solution to a particular environment. We evaluated \lang on a variety of applications, and found that its solutions were near optimal and performed comparably to hand-optimized configurations.


\label{endOfBody}



\bibliographystyle{plain}
\bibliography{references}

\appendix
\section{Preprocessing Heuristics}
\label{app:heuristics}
All three of our heuristics operate by attempting to assign each action in the Lucid program to a stage of the switch's pipeline. The primary distinction between the heuristics is the types of resources they account for during placement. Our simplest heuristic, dataflow graph, only accounts for dependencies between actions (two actions cannot be in the same stage if one depends on the output of the other). Our next heuristic, greedy layout, additionally considers the layout of memory, hash units, array accesses, and ALU usage (for example, we cannot have multiple concurrent accesses to the same array). 
Our final heuristic is to run a partial compilation -- rather than fully compiling to the switch, we instead compile Lucid to P4. This is much faster than a full compilation, and additionally considers resource limits on physical tables in the pipeline (such as match column width, maximum table size, and number of actions per stage). The only constraints that we encountered which were not modeled by the Lucid compiler are packet header vector (PHV) clustering constraints -- each packet header or metadata variable in a program must be placed into a specific PHV cluster, and each cluster has a fixed number of ALUs in each pipeline stage. In our experience,
it was possible to run afoul of PHV constraints in sufficiently complicated programs, but these violations were unaffected by choice of parameter values. Our preliminary implementations of 6/10 applications failed to compile with \emph{any} configuration due to PHV constraints, but once we adjusted the programs to accommodate for the constraints, we did not run into PHV constraint violations for any configurations.


\section{Comparison to Hand-Optimized Code}
\label{app:hand_opt}
We also compared \lang solutions to hand-optimized solutions for three of our applications: Fridge, Conquest, and Starflow. \lang's solutions performed reasonably close to the hand-optimized solutions for all 3 applications. We describe each application in more detail below.

\paragraph{Fridge (Unbiased RTT)} 
The Fridge\cite{fridge} data structure is used to collect RTT samples in the data plane by storing requests and matching them with the corresponding response, without sampling bias against large RTTs. Each request is added to the data structure with probability $p$, and once a request is in the structure, it can be removed either upon receipt of the response, or if a new response overwrites it due to a hash collision.


The value of $p$ is the primary parameter to be optimized. If $p$ is too small, requests are less likely to be added to the structure, and the program will not produce enough RTT samples. Conversely, if $p$ is too large, requests are more likely to be overwritten before their responses arrive. 

In general, the objective function that Fridge seeks to minimize is the difference between the distribution of sampled RTTs and the distribution of all RTTs. We implemented the same error function in \lang as was used in the original evaluation of Fridge~\cite{fridge}: maximum percentile error, or the maximum error of the sampled distribution for percentiles $\in [5\%,95\%]$. 

In the hand-tuned program, the authors achieved an error of 25\%, and our optimized program, found using Bayesian search, achieved a maximum delay estimation error of 28\%. 
The Fridge authors found that they could achieve nearly the same error with a wide range of $p$ values. In our workloads, \lang also found that $p$ had a negligible effect on error as long as it is greater than $2^{-12}$ ($0.0002$). Going outside of that bound for the chosen fridge size increased the error to over 100\%. 

\paragraph{Conquest.} 
Conquest~\cite{conquest} aims to identify flows that are making a significant contribution to queue build-up, during some time window $T$. It maintains several sketches as ``snapshots'' of the queue length for $T$. During a time window, the program cleans one sketch, writes to one sketch, and reads a flow's queue length estimates from the rest.

Conquest has three parameters that can impact its performance: the number of sketches and the rows and columns in each sketch. These parameters are challenging to tune because the choice of one affects the others. If the number of columns is too large, it reduces the number of rows that will fit on the target, and the sketch may not be fully cleaned before rotating. Conversely, too many rows requires less columns and smaller sketches. As a sketch gets smaller, it becomes less accurate.

The objective of Conquest is to identify the packets responsible for queue build-up as accurately as possible. For comparison with the original evaluation, we quantify accuracy using the F-score\footnote{Specifically, the cost is 1 minus the F-score}, which depends on both precision and recall.

The original evaluation of Conquest found that it could achieve both precision and recall greater than 90\%, i.e., an F-score $>$90\%. \lang found a comparable configuration with an F-score of 92\% (precision of 97\% and recall of 87\%). The \lang optimizer used the Bayesian search strategy, and the configuration was found after 9 iterations.

The choice of metric used for cost affects the configuration chosen by the optimizer. F-score incorporates both precision and recall. A configuration with lower precision has more false positives, and a lower recall means more false negatives. Some applications may be more tolerant to false negatives, and others may prefer false positives. We can tailor the objective function based on an application's preference. 

To minimize false positives, we can optimize for precision. This will result in a larger sketch, that keeps more accurate counts for each flow. On the other hand, we can optimize for recall to minimize false negatives. This produces a configuration with a smaller sketch, which will result in more flows being identified as significant contributors. In other words, more true positives, at the cost of more false positives as well. 


\paragraph{Starflow.} 
Starflow~\cite{starflow} is a telemetry system that partitions query processing between the data plane and software. The switch selects and groups per-packet records, which are sent to software for flow-level analytics (e.g., classifying traffic, identifying microbursts). Packet records are stored within buffers on the switch, and are evicted to software when their buffer is filled, no buffer is available, or there is a collision. There are two kinds of buffers, whose sizes must be configured at compile time: a ``narrow'' buffer which tracks many small flows, and a ``wide'' buffer for tracking a few large flows.

The most important performance metric for Starflow is its eviction ratio: the ratio of flushed cache records to packets. A lower eviction ratio is preferable because it means that more packets are being covered by each record that the server must process,  saving both bandwidth and processing time at server. 

The original, hand-optimized P4 code achieved an eviction ratio between 7.1\% and 25\%, depending on the size of the cache and the workload. The \lang optimizer achieved an eviction ratio of 15\%, well within the performance range of the original program. In other words, 15 out of every 100 packets are recirculated to evict a record from the cache.
The best compiling configuration was found after 7 (out of 85) iterations (1.5 min) of simulated annealing. 
We found that both the sizes of the narrow and wide buffers impacted the eviction ratio. Our optimizer found, for our representative traffic trace, that a narrow cache smaller than 1024 slots and a wide cache smaller than 8192 slots resulted in an eviction ratio greater than 40\%, with fixed wide and narrow caches, respectively.

\end{document}